\documentstyle[11pt,paspconf,epsf,psfig,twoside]{article}

\def\etal{{et\,al. }}
\def\msun{M$_{\odot}$}

\def\it{\sl}
\def\degs{\ifmmode ^{\circ}\else$^{\circ}$\fi}
\def\amin{\ifmmode ^{\prime}\else$^{\prime}$\fi}
\def\asec{\ifmmode ^{\prime\prime}\else$^{\prime\prime}$\fi}
\def\fdg{\hbox{$.\!\!^\circ$}}          % Fractions of degrees
\newbox\grsign \setbox\grsign=\hbox{$>$}
\newdimen\grdimen \grdimen=\ht\grsign
\newbox\laxbox \newbox\gaxbox
\setbox\gaxbox=\hbox{\raise.5ex\hbox{$>$}\llap
     {\lower.5ex\hbox{$\sim$}}}\ht1=\grdimen\dp1=0pt
\setbox\laxbox=\hbox{\raise.5ex\hbox{$<$}\llap
     {\lower.5ex\hbox{$\sim$}}}\ht2=\grdimen\dp2=0pt
\def\gax{$\mathrel{\copy\gaxbox}$}

\def\rxj{RX\,J0719.2+6557}

\begin{document}

\title{High spectral and time resolution observations of the eclipsing polar 
 RX\,J0719.2+6557}

\author{G.\,H. Tovmassian$^{1}$, P. Szkody$^{2}$,  J. Greiner$^{3}$,
  S. Vrielmann$^{4}$, P. Kroll$^{5}$,\\ S. Howell$^{6}$, R. Saxton$^{6}$, 
  D.Ciardi$^{6}$, P.A. Mason$^{7}$ and N.C. Hastings$^{2}$}
\affil{$^{1}$OAN, Instituto de Astronom\'{\i}a,UNAM, M\'{e}xico\\
$^{2}$Department of Astronomy, University of Washington, Seattle, USA \\
$^{3}$Astrophysical Institute Potsdam, Potsdam, Germany\\
$^{4}$Dept. of Astronomy, University of Cape Town, South Africa\\
$^{5}$Sternwarte Sonneberg, 96515 Sonneberg, Germany\\
$^{6}$University of Wyoming, Laramie, Wyoming, USA\\
$^{7}$Department of Astronomy, NMSU, Las Cruces, New Mexico, USA}
%
% Footnotes to the authors or title don't work well with this method --
% acknowledgments of this sort should go in the acknowledgments at the
% end. 

% The abstract is entered in a LaTeX "environment", designated with paired
% \begin{abstract} -- \end{abstract} commands.  Other environments are
% identified by the name in the curly braces.

\begin{abstract}
We present phase-resolved spectral and multicolor CCD-photometric observations 
of the  eclipsing polar RX\,J0719.2+6557 obtained with relatively high time 
($\approx$600 sec/15 sec) and spectral (2.1 \AA) resolution when the
system was in a high accretion state. The trailed spectrograms clearly reveal
the presence of three different line components with different width
and radial velocity variation. The Balmer emission
lines, as well as the higher excitation He\,{\sc ii} line, contain significant 
contributions from the X/UV-illuminated hemisphere
of the companion star. We were able to resolve all three components by line 
deblending, and by means of  Doppler tomography were able to unambiguously 
identify the emission components  with the secondary star,  the ballistic
part of the accretion stream  and the magnetically
funnelled part of the stream. The light curves and eclipse profiles provide 
additional information about the system geometry. 
\end{abstract}

% Keywords can be included, but they are not printed in the hardcopy.
% These are useful for you to point out words you want in the index

\section{Introduction}
The X-ray source \rxj\, was discovered during the ROSAT All Sky Survey and 
identified as an eclipsing polar by Tovmassian \etal(1997)(T97). 
The authors found that the system undergoes deep eclipses (3 mag), 
with flat light curve out of eclipse in the blue and  with pronounced 
sinusoidal variations in red light.
The orbital period estimated at $P_{\rm  orb}=0.068207$ day based on the 
eclipse occurrence, coincided with radial velocity variations obtained from  
the spectroscopy. T97 showed that the line profiles are multi-component and  
were able to identify two components with difficulty.

Here we report on new  observations of \rxj\, with improved spectral and time 
characteristics that allow us to conduct a detailed study of the various 
components of this eclipsing polar. 

\section {Data acquisition and reduction}

We obtained spectra of \rxj\, during two half nights with the 3.5m ARC 
telescope at Apache Point Observatory using the DIS spectrograph set in 
two wavelength regions $4200-5100\rm \AA$ and $7900-8900 \rm \AA$ with 
spectral resolution of $1.3 \rm {\AA/pix}$, corresponding to overall FWHM 
resolution $2.4 \rm \AA$. We also acquired fast photometry at APO using 
SpiCam in three broad band filters (B\,V\,R), covering one orbit in each band.
 We observed the object in the IR J\,H\,K filters simultaneously with the 
spectroscopy during poor weather conditions in Wyoming. We also
collected more unfiltered (red)  light CCD-photometry from Sonneberg and 
Red Buttes Observatory (RBO).

\section {Analysis and results}
\subsection {Orbital period and eclipse phase}

Based on the new photometric data combined with eclipse moments 
reported in (T97) we refined the orbital period of \rxj. 
We should note here that this ephemeris refers to the eclipse phasing 
or photometric phase. Later we will show that the  spectral phase differs from 
the photometric phase.

\subsection {Irradiated secondary} 

\begin{figure}
\psfig{file=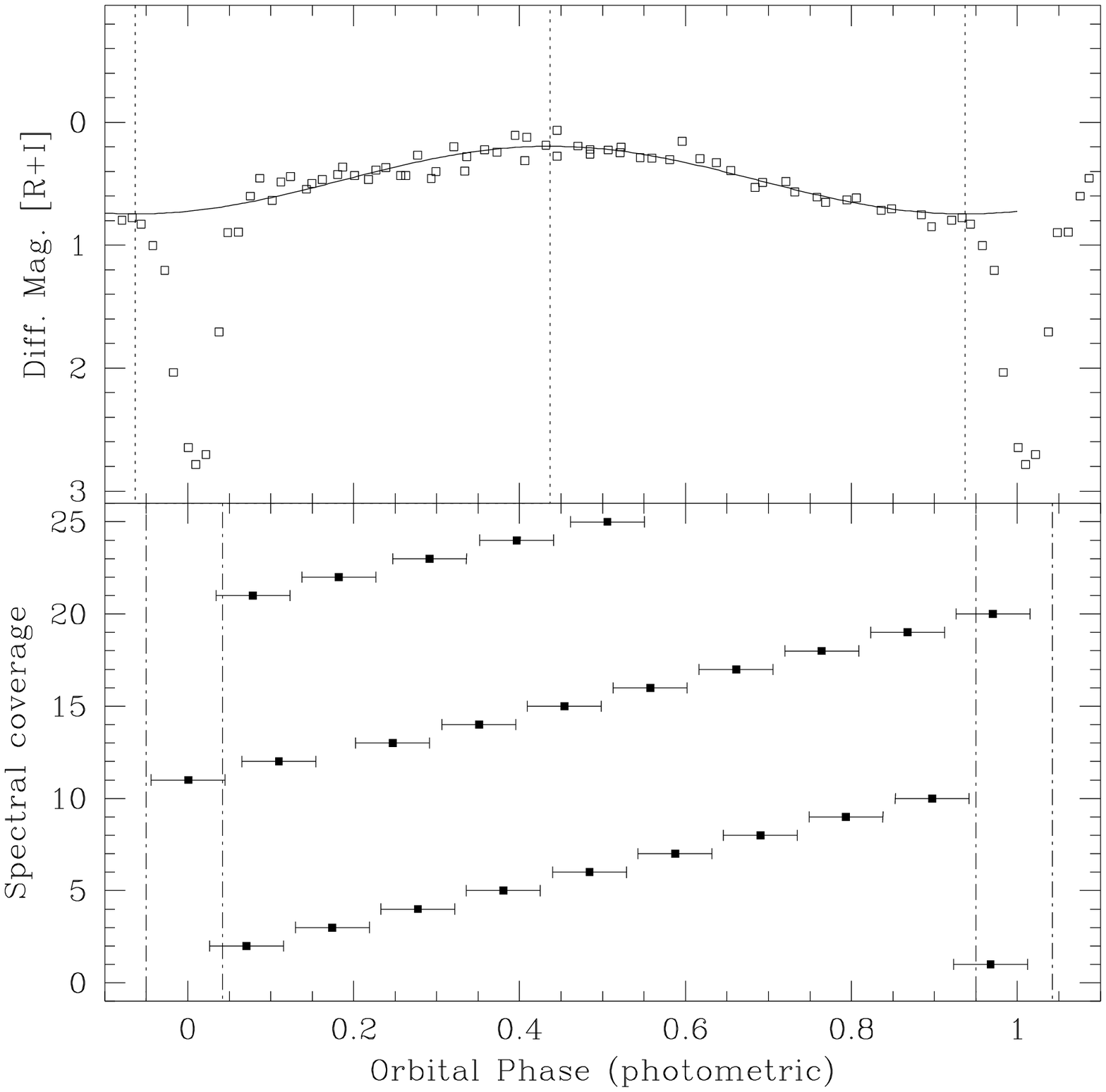,width=50mm}
\vspace*{-5.3cm}\hspace{5.cm}
\psfig{file=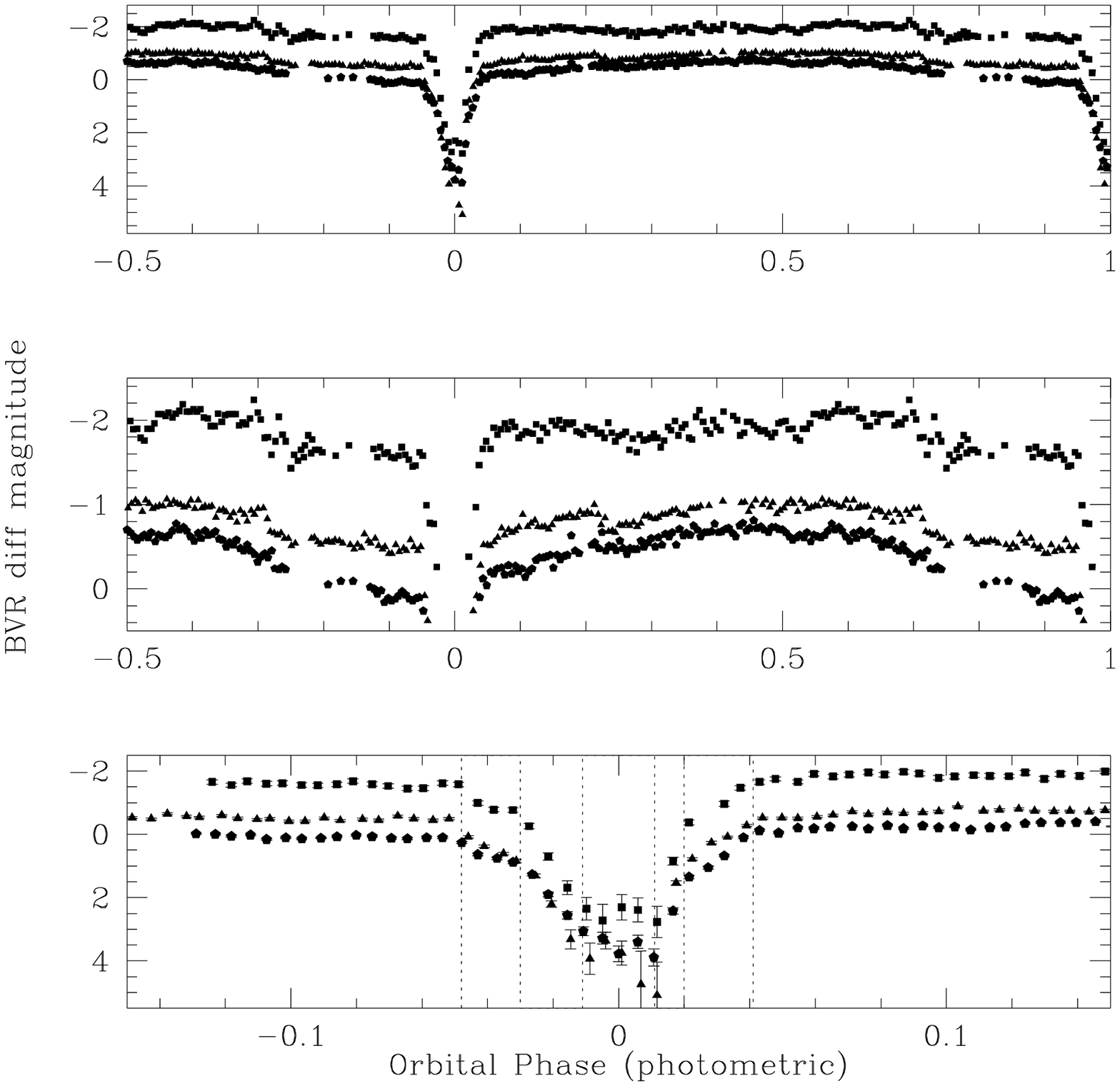,width=82mm}
\parbox[t]{6.1cm}{\vspace*{-3.1cm}
\caption{Left: R light curve and fit by a $sin$ curve 
  (out of eclipse, top) and the spectroscopic coverage (bottom). 
  The eclipse and the  minima of the $sin$ curve}}
%\vspace*{-1.cm}
\parbox[t]{13.cm}{\vspace*{-.56cm}\hspace{.66cm} are shown by vertical lines. 
   Right: BVR light curves at different scales.}
\label{fg1}
\vspace*{-0.35cm}
\end{figure}

We used the optical photometric data obtained without filter at RBO with CCD 
sensitive in the far red (corresponding to R+I) along with B\&R light curves 
obtained at the 3.5m ARC telescope for a study of periodic variations outside 
the eclipse.  After eliminating the eclipse points, we fitted the 
light curves with a simple sine function. In the case of the red unfiltered 
light curve we just dismissed the points corresponding to the eclipse. For the
ARC data we normalized  the depth of eclipse in B to the one in R and 
subtracted the former from latter. The red light curve  obtained at RBO and 
the fitted curve are shown in Fig. \ref{fg1}.  The minima in the R band
occur  at $\delta\phi_{\rm orb}=0.06$  prior to the 
eclipse. An identical result was  obtained for the APO data.
The modulation is single humped relative to the orbital period. We presume 
that this modulation in red light is due to the irradiated side of the 
secondary component. The phase lag between the minimum of the single 
humped curve and the eclipse has two possible explanations, which might both 
contribute to the  large amount of the observed lag. One explanation is 
that the occulting source is not the white dwarf, but the bright spot on 
the accretion stream in the coupling region. Another possible cause of the 
phase lag is uneven irradiation of the leading side of the secondary due to 
its obscuration  by a stream leaving the L1 point. 

\subsection {Eclipse light curves}

Fig. \ref{fg1} (right panel) shows three eclipse light curves  in different 
broad band 
filter B, V \& R. They are presented on three panels with different scales 
to distinguish details at different orbital phases. On the lower panel one 
can see that the eclipse actually consists of two parts. The first section 
is relatively broad and shallow, while the second is narrower and deeper. 
The $\approx 4$ mag depth of eclipse in B is evidence that the compact 
structure that is occulted is a major source of light. The errors at the 
bottom of eclipse are too big to be able to make more precise estimates. 
The various steps of eclipse are marked by vertical lines in the figure. 
The total eclipse lasts about 505 sec. Meanwhile the deep eclipse duration 
at half depth is $245\pm20$ sec, which corresponds to half width at half 
depth $\phi_{1/2}=7\fdg5\pm2\fdg0$ as defined by Bailey (1990). 

The eclipse profile is similar to that of 1H\,1752+081 possibly
containing an accretion disk (Silber \etal 1994), although there 
are contradictions in photometry and spectroscopy with that explanation. 
At the same time,  Barwig \etal (1994) interpret this eclipse profile 
as an eclipse of the white dwarf followed immediately by 
the accretion stream, i.e. the white dwarf is being eclipsed 
first (step 1), then the accretion stream (step 2). Step 3 corresponds to the 
egress of white dwarf and finally egresses the stream (step 4). It makes sense
in the case of 1H\,1752+081 where the depth of both eclipse phases (1 \& 2) 
are about the same, and one can pair 1st with 3rd  (2nd with 4th) on the basis 
of their steepness. However, in case of \rxj\, this scenario is not 
applicable, since the initial occultation (step 1) is very shallow as well 
as the final egress (step 4) in comparison to the central dip. One remote 
possibility would be to suppose that the first step 
corresponds to the ingress of unspotted part of the white dwarf, while the 
second phase is the mixed ingress of spot on the second half of the  white 
dwarf and stream, with white dwarf egress resulting in step 3 and finally 
stream (step 4). To test this scenario higher accuracy 
and faster photometry at the bottom of the eclipse would be needed
to prove if it is either flat as 
it seems in the B light curve or structured as it looks in V and R data. 
Until then, taking into account the above mentioned phase lag between eclipse 
and minimum in red light and possible variability of the eclipse profile, we 
tend to think that a more plausible cause of the eclipse is the occultation of 
the stream with the bright spot elevated some distance over the white dwarf 
surface. 

\subsection {Spectroscopy}

\begin{figure}
\psfig{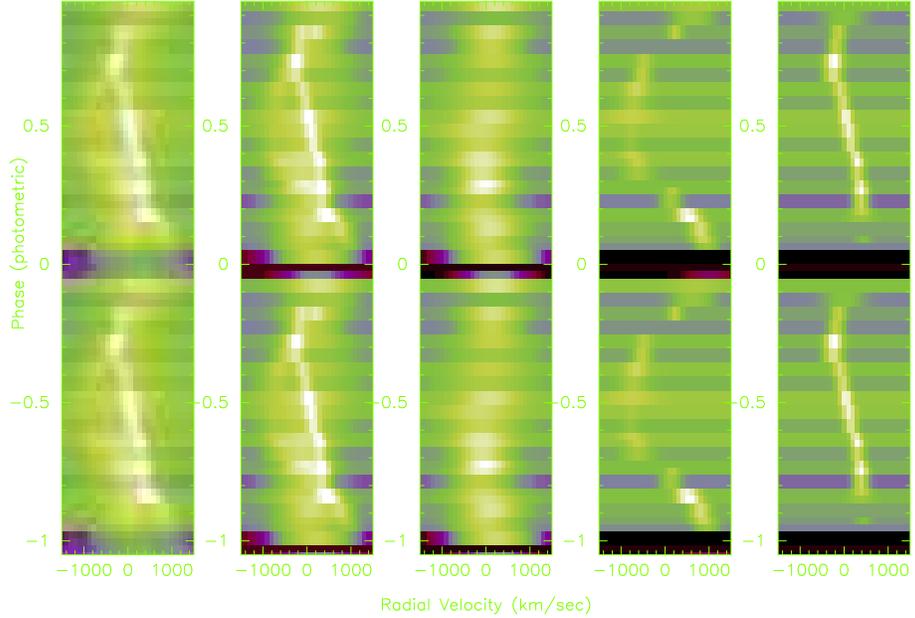}
\psrotatefirst
\vspace{-0.35cm}
\caption{Trailed He\,{\sc ii} spectra of \rxj: as observed (left), the 
  computed composite 
  of three gaussians (second from left) and the three gaussians separately.}
   \label{fg2}
\end{figure}

We used the three brightest lines in our spectral range H$\beta$, 
H$\gamma$ and He\,{\sc ii} for our analysis.  The line profiles are very 
complex. They look similar, but the Balmer lines are more blury than 
He\,{\sc ii} and the components seem less resolved.
Hence, in order to study the components of the emission lines, we started 
with the He\,{\sc ii} line. We stacked the spectra into a bidimensional image 
to construct  trailed spectra (Fig. \ref{fg2}). They were placed in order 
of orbital phases (calculated according the photometric ephemeris). 
Each line corresponds to approximately 0.05 orbital phase.   
In the trailed spectra the so called narrow component is easily distinguished.
 It is more pronounced  at phases 0.15--0.85 being the brightest 
feature there. The other noticable feature is a high velocity component 
(HVC) which is shaping the edges of the emission line around phases 
0.0 and 0.5. 
%In Fig.\ref{fg2} the trailed spectra of He\,{\sc ii} line is 
%presented on the left panel. Next to it the sum of three gaussians is 
%displayed, with each of them presented separately in the subsequent panels.

\begin{figure}
\begin{minipage}{55mm}
\psfig{file=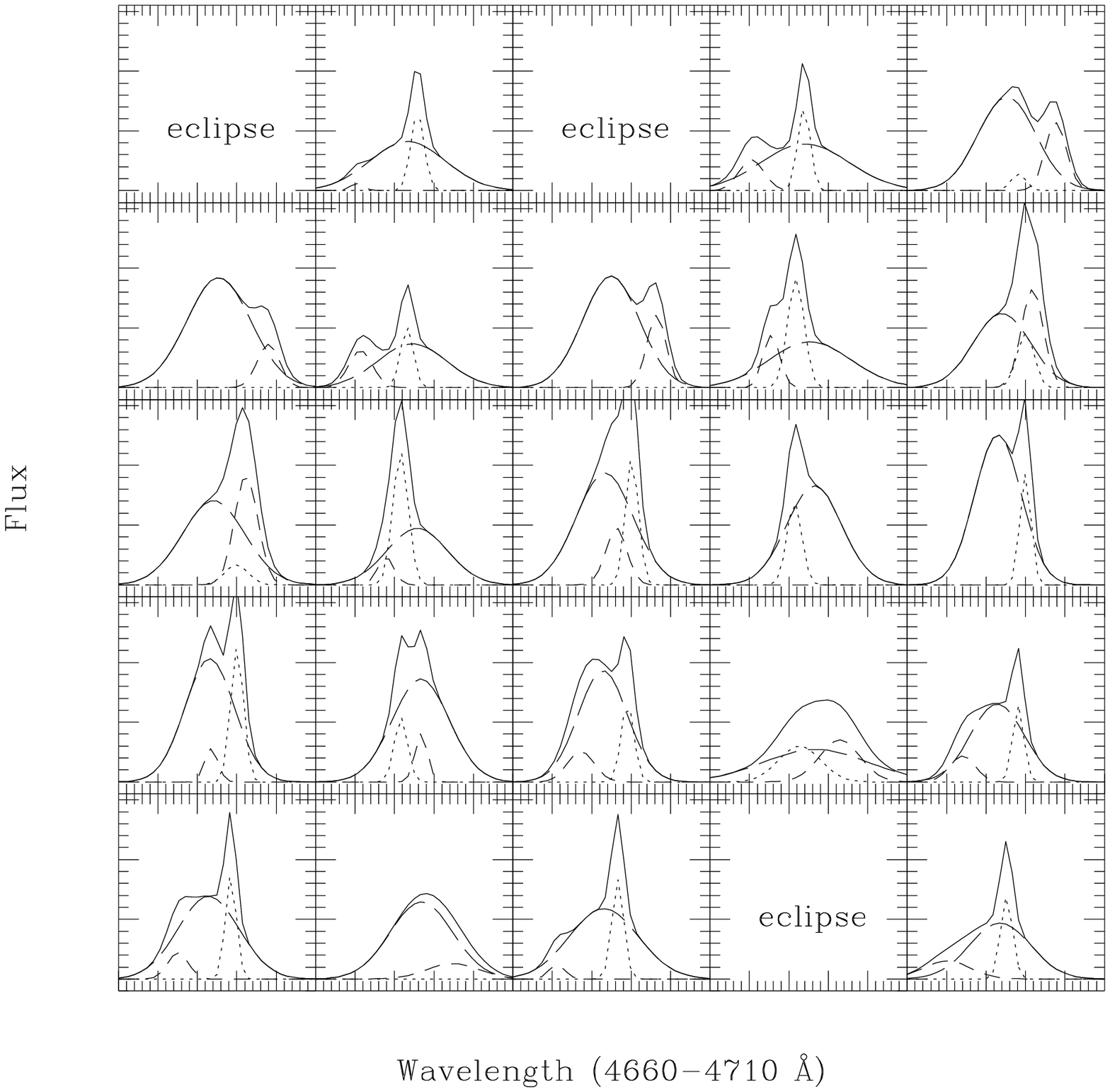,width=65mm}
\end{minipage}
%\hspace*{0.1cm}
\begin{minipage}{55mm}
\psfig{file=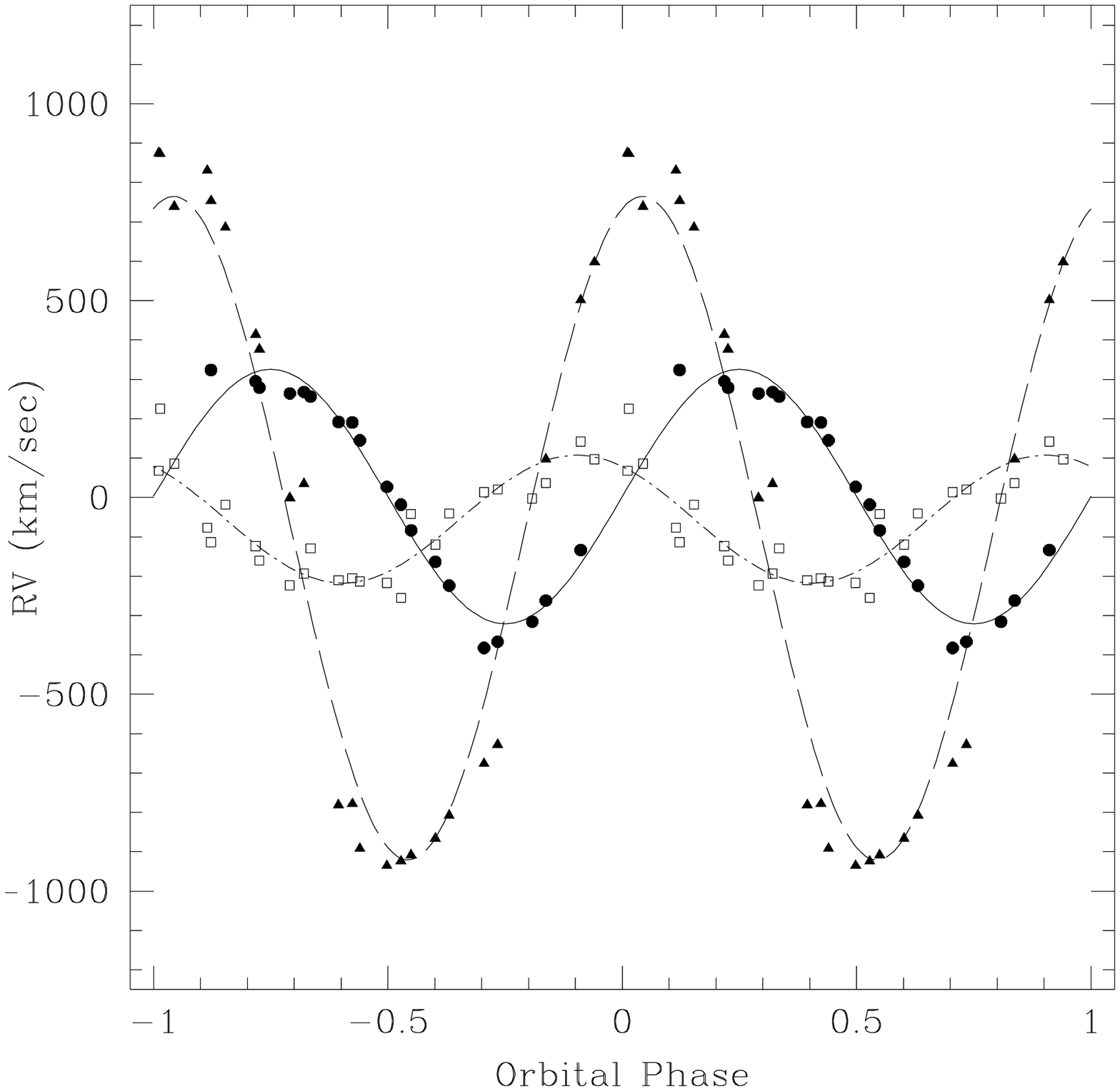,width=67mm}
\end{minipage}
\vspace{-0.35cm}
\caption{Left: Profiles of the He\,{\sc ii} line and deblending  components
   (dashed lines). 
   The solid line is the sum of all three components. Right: The radial 
   velocities of the three components and $sin$ fits.}
\label{fg3}
\end{figure}

The common procedures for line measurements, such as gaussian deconvolution 
for example,  do not work in this case. Thus, we measured the above mentioned 
component centers ``manually" on the zoomed trailed spectra image. These gave 
us enough points to fit sine curves to both of them with the photometric 
period and to estimate the central wavelenghts of both 
components at each phase. We used these interpolated line centers in the line 
deblending procedure available in IRAF to define the central wavelengths of 
the components more precisely. We did another iteration  supplying the fitted 
values to the deblending procedure and found  that two components  can 
not account for the entire line profile. We subtracted both components from 
the actual line profile and measured the center of the remaining broad 
component. It also displayed wavelike variation appropriate to the orbital 
period of the system. We did yet another iteration by supplying the deblending
 procedure with three components before we were satisfied with the results.  
At eclipse and phases immediately after or before, we failed to identify all 
three components, and at some phases it was difficult  to distinguish between 
them. Nevertheless the resulting profiles and corresponding  radial velocity 
curves are quite reasonable (Fig. \ref{fg3}). 
%The profiles of the components of He\,{\sc ii} 
%line are presented in the left panel of Fig. \ref{fg3} by dashed lines. 
%The solid line is the  sum of them.  
The radial velocity curves of 
three components are also presented in Fig. \ref{fg3} (right panel). 
Among other interesting 
features to be seen in the radial velocity phasing, the most striking is 
the time lag between the zero velocity crossing of secondary (conjunction) 
and eclipse, which happens $\delta\phi=0.06$ phase later. That coincides 
exactly with the result drawn from photometry. The similar results were 
obtained from hydrogen lines analysis.

\section {Some constraints}

\begin{figure}
\psfig{figure=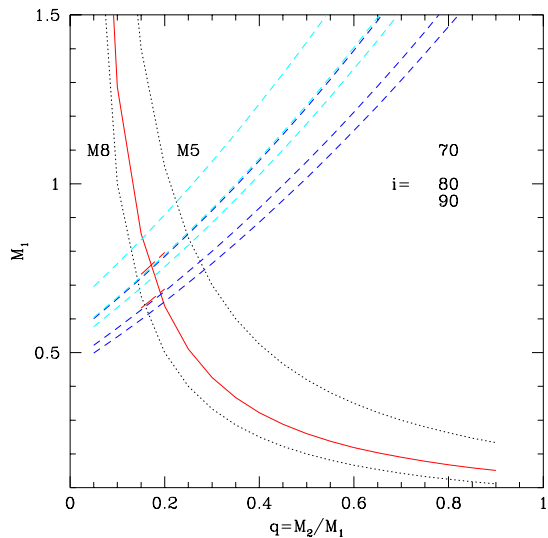,width=7.5cm}
\hfill
%\parbox[t]{6.9cm}{\vspace*{-7.8cm}
\parbox[t]{6.9cm}{\vspace*{-6.cm}
\caption{Accretor mass versus mass ratio. Dashed lines mark inclinations of
  $i$=90\deg, 80\deg\ and 70\deg\ (from bottom to top) with the two values
  of K2=400 and 450 km/s each. Two M$_{\rm 1}$ values are marked for 
  different secondary masses depending on their spectral class (dotted lines)
  and the condition for a Roche lobe filling secondary is drawn as solid line.
  }}
   \label{fg4}
  \vspace*{-.35cm}
\end{figure}

In summary,  we have the following observational data for consideration of the 
system parameters: 
(i) the orbital period $P_{orb}=5893.13$ sec,
(ii) the radial velocity amplitude ${K_2}^{'}$=323 km/sec for He\,{\sc ii} and 
379 km/sec for composite $H\beta$/$H\gamma$ data
(iii) the eclipse half width at half depth $\phi_{1/2}=7\fdg5\pm2\fdg0$ (as 
defined by Bailey 1990) and
(iv) $i$ \gax 75\deg\ due to eclipses. 
Now, if we make a number of assumptions, we can estimate 
different parameters of the system. The basic assumptions common for most 
CVs are that the secondary is a pre-main sequence star filling its Roche 
lobe and obeying the mass-radius relation for main-sequence stars.
Assuming in addition that the RV of the narrow line component reflects
the  motion of  the irradiated secondary, then after applying a 
reasonable correction for the
center of mass of the secondary, we get a range of 400 to 450 km/sec.  
Taking into account  the inclination, period and  
velocity amplitude we can constrain the mass of the primary WD. 
The resulting diagram of the $M_1$ dependence from $\rm {q=M_2/M_1}$ is 
plotted in Fig. \ref{fg4}.
Combining the relation between 
the binary component masses and the Roche lobe size
as suggested by Iben \& Tutukov (1984), with the mass-radius relation of 
low-mass main-sequence stars  we calculate the condition for a Roche lobe 
filling secondary. The resulting q range of 0.15--0.3 is
in agreement with the expected M~4--5.5 spectral type for the secondary
according to the orbital period to spectral type relation.

We can further constrain the inclination and mass ratio by using the observed 
eclipse width if we assume that the occulted source is the white dwarf.
 If the companion in \rxj\, fills its Roche lobe, the eclipse
half width at half depth  of about  7\fdg5
defines a unique relation between the mass ratio q
and the inclination of the system with respect to the line of sight (Chanan
\etal 1976). This relation crosses the line of the Roche lobe filling
condition at i=79$\pm$3\deg. If the secondary is an unevolved main-sequence
star just filling its Roche lobe, then this inclination implies the following
solution: a white dwarf mass of $\rm {M_1=0.75}\pm0.1$\msun\, and a companion
mass of $\rm {M_2=0.13}\pm0.05$\msun\, (q=0.175). In the case of an M5 
companion (implying an evolved state to allow Roche lobe filling) 
$\rm {M_1=0.85}\pm0.1$\msun\, and $\rm {M_2=0.21}\pm0.05$\msun\, (q=0.25)
is also a viable solution.

% Now comes the reference list.  Since we typed out the citations ourselves,
% the reference list is enclosed in a "references" environment.  Each
% new reference begins with a \reference command which sets up the proper
% indentation.  Typography that may be required in the reference list by
% the editorial staff must be included by the author.
%
% Observe the "standard" order for bibliographic material: author name(s),
% publication year, journal name, volume, and page number for articles.
% Some journal names are available as macros; see the WGAS markup
% instructions for a listing of which ones have been "macro-ized".
% Note the use of curly braces to delimit the font changes: it is essential
% that this be done to limit the scope of the font declaration.
%
% There is no need to engage in any other typographic manipulation.

\end{document}